\documentstyle[12pt]{article}
\def\newpic#1{%
   \def\emline##1##2##3##4##5##6{%
      \put(##1,##2){\special{em:point #1##3}}%
      \put(##4,##5){\special{em:point #1##6}}%
      \special{em:line #1##3,#1##6}}}
\newpic{}
\newcommand{\be}{\begin{equation}}
\newcommand{\ee}{\end{equation}}
\newcommand{\ba}{\begin{eqnarray}}
\newcommand{\ea}{\end{eqnarray}}
\newcommand{\baa}{\begin{eqnarray*}}
\newcommand{\eaa}{\end{eqnarray*}}
\newcommand{\bb}{\begin{thebibliography}}
\newcommand{\eb}{\end{thebibliography}}
\newcommand{\ci}[1]{\cite{#1}}

\newcommand{\lab}[1]{\label{#1}}
\newcommand{\re}[1]{(\ref{#1})}
\newcommand{\nn}{$1/n$}
\newcommand{\tr}{{\rm tr\,{\bf 1}}}

\newcounter{hran}

\newcommand{\al}{\ifmmode\alpha\else$\alpha$\fi}
\newcommand{\psb}{\ifmmode\overline{\psi}\else$\overline{\psi}$\fi}
\newcommand{\pl}{\ifmmode\partial\else$\partial$\fi}
\newcommand{\plh}{\ifmmode\widehat{\pl}\else$\widehat{\pl}$\fi}
\newcommand{\gm}{\ifmmode\gamma\else$\gamma$\fi}
\newcommand{\Gm}{\ifmmode\Gamma\else$\Gamma$\fi}
\newcommand{\veps}{\ifmmode\varepsilon\else$\varepsilon$\fi}
\newcommand{\eps}{\ifmmode\epsilon\else$\epsilon$\fi}
\newcommand{\bt}{\ifmmode\beta\else$\beta$\fi}
\newcommand{\gps}{\ifmmode\gm_{\psi}\else$\gm_{\psi}$\fi}
\newcommand{\gpsb}{\ifmmode\gm_{\psb}\else$\gm_{\psb}$\fi}
\newcommand{\sg}{\ifmmode\sigma\else$\sigma$\fi}
\newcommand{\dl}{\ifmmode\delta\else$\delta$\fi}
\newcommand{\Dl}{\ifmmode\Delta\else$\Delta$\fi}
\newcommand{\Dps}{\ifmmode\Dl_{\psi}\else$\Dl_{\psi}$\fi}
\newcommand{\dps}{\ifmmode d_{\psi}\else$d_{\psi}$\fi}
\newcommand{\Dpsb}{\ifmmode\Dl_{\psb}\else$\Dl_{\psb}$\fi}
\newcommand{\dpsb}{\ifmmode d_{\psb}\else$d_{\psb}$\fi}
\newcommand{\Dsg}{\ifmmode\Dl_{\sg}\else$\Dl_{\sg}$\fi}
\newcommand{\om}{\ifmmode\omega\else$\omega$\fi}
\newcommand{\et}{\ifmmode\eta\else$\eta$\fi}
\newcommand{\vph}{\ifmmode\varphi\else$\varphi$\fi}
\newcommand{\LL}[1]{\ifmmode{\cal L}_{#1}\else${\cal L}_{#1}$\fi}
\renewcommand{\v}{\ifmmode{\cal V}\else${\cal V}$\fi}

\newcommand{\th}[2]{\ifmmode\theta^{#1}_{#2}\else$\theta^{#1}_{#2}$\fi}
\newcommand{\Ph}[1]{\ifmmode\Phi^{#1}\else$\Phi^{#1}$\fi}
\newcommand{\J}[2]{\ifmmode J^{#1}_{#2}\else$J^{#1}_{#2}$\fi}
\newcommand{\NN}[1]{\ifmmode N^{#1} \else${N}^{#1}$\fi}
\newcommand{\T}[1]{\ifmmode T^{#1}\else$T^{#1}$\fi}
\renewcommand{\S}[1]{\ifmmode S_{#1}(\vph)\else$S_{#1}(\vph)$\fi}
\newcommand{\ld}[1]{\ifmmode\lambda_{#1}\else$lambda_{#1}$\fi}
\newcommand{\DA}[1]{\ifmmode {\cal D}_A^{#1}\else${\cal D}_A^{#1}$\fi}
\newcommand{\OO}[1]{\ifmmode {\cal O}_{#1}\else${\cal O}_{#1}$\fi}
\title{On Calculation of $2+\veps$ RG Functions in the Gross Neveu Model
from Large $N$ Expansions of Critical Exponents.}
\author{
{Kivel N.A.\ ,\ \ Stepanenko A.S.\thanks{E-mail address:
stepan at amoco.saclay.cea.fr}}\\
{\small
\it Theoretical Department, St.-Petersburg Nuclear Physics
Institute,}\\
{\small
\it Gatchina, St.-Petersburg, 188350, Russian Federation}\\
{Vasil'ev A.N.\thanks{This work was supported by the Grant of the American
Physical Society}\thanks{E-mail address: ioffe
at onti.phys.lgu.spb.su}}\\ {\small \it Department of
Theoretical Physics, St.-Petersburg University,}\\ {\small
\it Ul'yanovskaya 1, Staryi Petergof, 198904, St.-Petersburg,
Russian Federation}}
\date{  }

\begin{document}
\setcounter{page}{0}
\maketitle
\vskip -9.5cm
\rightline{preprint PNPI--1910}
\vskip 9.5cm
\thispagestyle{empty}
\begin{abstract}
        Using knowledge of the explicit $n$ dependence of the RG
        functions and expressions of critical exponents in the
	framework of large $N$ expansion in the Gross Neveu model we
        derive RG functions in 4- and 5-loop approximation.
\end{abstract}
\newpage

\section{Introduction.}
        It is well known that the structure of renormalized field theory is
        determined by the renormalization group (RG) equations. One
        considerable problem then is the calculation of renormalization
        group functions.  Usually one can compute only a few orders in
        perturbation theory.  Hence, we do not have a complete knowledge of
        these functions.  Other methods, such as the large $N$ expansion,
        usually allow us to compute critical exponents at leading order only
        because calculations become too complicated at higher orders.

        To overcome these difficulties other methods have been developed.
        One such approach uses critical conformal
	invariance at the fixed point of renormalization group for
	calculation of critical exponents, which are values of RG
	functions at fixed point. With this method it is possible to
	derive the critical exponents at order $1/N^3$. Using then
	the exact $N$ dependence of RG functions at each order in
	perturbation theory one can reconstruct them at some
	high-loop approximation.

        In this paper we derive five loop approximation for RG functions
        (up to few unknown coefficients) in the Gross Neveu and the
        non-linear $\sigma$ models.

\section{Preliminaries.}

        The $U(N)$--symmetric massive Gross Neveu (GN) model describes a
        system of $N$ $D$--dimensional Dirac spinors, its non-renormalized
        action being
\be
        S = \int\!{\rm d}^Dx\ [\psb_a(\plh+m_0)\psi_a+g_0(\psb_a\psi_a)^2
        /2]
\lab{1}
\ee
        $a=1,\dots,N$ and summation over repeated indices is implied
        (in what follows we consider the theory in Euclidean space and
        functional distribution is denoted by $\exp S$ without the minus in
        the exponent). It is believed to be multiplicatively
        renormalizable in the framework of $2+\veps$ expansion and its
        renormalized action in dimension  $D=2+2\veps$ is
\be
        S_R=\int\!{\rm d}^Dx\ [Z_0m^2M^{2\veps}+\psb_a(Z_1\plh+Z_2m)\psi_a+
        gM^{-2\veps}Z_3(\psb_a\psi_a)^2/2] \ ,
\lab{2}
\ee
        where $g$ is a dimensionless renormalized coupling constant, $m$ is
        a renormalized mass, $M$ is a renormalization mass and
        $Z_{0,1,2,3}$ are renormalization constants. The counterterms of
        model \re{2} (without vacuum one) are generated by the standard
        multiplicative renormalization procedure
\be
        \psi\rightarrow Z_\psi\psi\ ,\ \ m_0=mZ_m\ ,\ \
        g_0=gZ_gM^{-2\veps}\ ,
\lab{1a}
\ee
\be
        Z_1=Z_\psi^2\ ,\ \ Z_2=Z_\psi^2Z_m\ ,\ \ Z_3=Z_\psi^4 Z_g\ .
\lab{2a}
\ee
        RG functions $\gamma_F$ (the anomalous dimensions of quantities
        $F$) and $\beta$ function are defined by the relations
\be
        \gamma_F=\tilde{\cal D}_\mu \ln Z_F\ (F=\psi,m,g)\ ,\ \
        \beta=\tilde{\cal D}_\mu g=(2\veps g-\gm_g)g\ ,\ \
        \tilde{\cal D}_\mu \equiv \mu\left.\frac{\rm d}{{\rm
        d}\mu}\right|_{g_0,m_0}\ .
\lab{2b}
\ee
        The $\beta$ function of model \re{2} has an UV fixed point
        $\beta(g_*)=0$ with $g_*\sim\veps$. Critical dimensions $\Dl_F=
        d_F+\gm_F^*$ (canonical + anomalous) of various quantities $F$
        (fields and parameters) in the framework of $2+\veps$ expansion are
        defined by the relations
\be
        \Dl_F=d_F+\gm_F^*\ (F=\psi,m)\ ,\ \ \Dl_\tau\equiv 1/\nu=-
        \beta^\prime(g_*)\ ,
\lab{2c}
\ee
        where $\gm^*_F\equiv \gm_F(g_*)$, $\Dl_\tau\equiv 1/\nu$  is the
        critical dimension of ``temperature'' $\tau=g-g_*$ and $d_F$ are
        the corresponding canonical dimensions: $d_\psi=D/2-1,\ d_m=1$.

        For practical calculations it is more convenient to use the
        following model:
\be
        S = \int\!{\rm d}^Dx\ [\psb_a(\plh+m_0)\psi_a-\sg^2/2g_{0}+
        \sg\psb_a\psi_a]\ .
\lab{13}
\ee
        The generalization of this model which is multiplicatively
        renormalizable in the framework of $2+\veps$ expansion is
\be
        \bar{S} = \int\!{\rm d}^Dx\ [\psb_a(\plh+\bar{m}_0)\psi_a-\sg^2/2+
        v_{0}\sg\psb_a\psi_a+g_{20}(\psb_a\psi_a)^2/2+h_0\sg]
\lab{10}
\ee
        with two independent bare charges $g_{20}$ and $g_{10}\equiv
        v_0^2$. Its renormalized form is
\ba
        \bar{S}_R&=&\int\!{\rm d}^Dx\ [\bar{Z}_0\bar{m}^2M^{2\veps}+
        \psb_a(\bar{Z}_1\plh+ \bar{Z}_2\bar{m})\psi_a- \bar{Z}_3\sg^2/2+
        \nonumber\\
        &&+\bar{Z}_4v_{1}M^{-\veps}\sg\psb_a\psi_a+
        \bar{Z}_5g_2M^{-2\veps}(\psb_a\psi_a)^2/2+
        \bar{Z}_6\bar{m}M^\veps\sg+h\sg]\ .
\lab{11}
\ea
        If we integrate model \re{11} over field \sg\ we obtain model
        \re{2} with parameters:
\be
        m=\bar{m}+hvM^{-\veps}\ ,\ \ g=g_1+g_2\ ,\ \ g_1\equiv v^2
\lab{11a}
\ee
        and additional constant ${\rm tr}\ln([\hat{\partial}+m]/
        [\hat{\partial}+\bar{m}])+(h^2/2)\int\!{\rm d}^Dx$ to the action.
        Comparing counterterms in the models \re{2} and \re{11} one can
        obtain the following relations for renormalization constants
\be
        \left.
        \begin{array}{l} \bar{Z}_1=Z_1\ ,\ \
        \bar{Z}_3^{-1}=1+2g_1(Z_0+n/8\pi\veps)\ ,\ \
        \bar{Z}_2=\bar{Z}_4=Z_2\bar{Z}_3\ ,\\
        v\bar{Z}_6=1-\bar{Z}_3=2g_1(\bar{Z}_0+n/8\pi\veps)\ ,\ \
        g_1\bar{Z}_3^{-1}\bar{Z}_4^2+g_2\bar{Z}_5=g\bar{Z}_3
        \end{array}
        \right\}
\lab{11b}
\ee
        where $g_1\equiv v^2$ and $n$ is defined by \re{4}.
        These relations allow to express 7 constants $\bar{Z}_i(g_1,g_2)$
        of model \re{11} through 4 constants $Z_i(g=g_1+g_2)$ of
        model \re{2}. It is more convenient to calculate the
        renormalization constants in the framework of massless
        ($\bar{m}=0,h=0$) one-coupling model like \re{11} with interaction
        $\sim\psb\psi\sg$ and additional four-fermion counterterm which
        breaks the multiplicative renormalizability.

        To generate a systematic $1/n$ expansion one usually uses the model
        \re{13} with a scalar field $\sg$.
        As have been shown in \ci{TMF} the massless version of \re{11}
        possesses critical conformal invariance at the $D$-dimensional
        fixed point of renormalization group and the exponent $\eta$ have
        been calculated up to order $1/n^3$ with the conformal
	bootstrap method.

        In what follows we shall consider \nn\ expansions in arbitrary
        dimension of space $D$ and the following notations will be
	used:
\be
        N\tr \equiv n\ , \ \ D \equiv 2\mu\ ,
\lab{4}
\ee
        where $\tr$ is a trace of the unit matrix in the space of
        $D$-dimensional spinors.
        In the framework of the \nn\ expansion for the
        model \re{13} the canonical dimensions of the fields $\vph
        \equiv \psi,\sg$ and parameter $\tau$ are
\be
	d_\psi=d_{\psb}=\mu-1/2\ ,\ \ d_\sg =1\ ,\ \ d_\tau = 2(\mu-1)\ .
\lab{5}
\ee
        The notation $d[F]\equiv d_F$ means canonical dimension of quantity
        $F$. Let us parametrize the critical dimensions $\Dl_F=d_F+\gm_F^*$
        (canonical $+$ anomalous) as follows:
\be
	\Dps=\dps+\eta/2\ ,\ \ \Dsg=d_\sg-\eta-2\Dl\ ,\ \ \Dl_\tau \equiv
        1/\nu \equiv 2\ld{} \ \ (\ld{0}=\mu-1)\ ,
\lab{6}
\ee
        where $\eta$ and $\nu$ are the usual notations for the critical
        exponents, the quantity $2\Dl$ is a critical dimension of the
        vertex $\psb\psi\sg$ in the model \re{13}. For this model
        the quantity $\Dl_\tau$ is connected with the critical
        dimension $\Dl[\sg^2]$ of a composite operator $F=\sg^2$
        by means of the following relation
\be
	\Dl_\tau \equiv 1/\nu \equiv 2\ld{} = 2\mu-\Dl[\sg^2] \ .
\lab{7}
\ee
        For the \nn\ expansion of any exponent $z$ its coefficients in
        powers of \nn\ are denoted by $z_k$:  $z=z_0+z_1/n+z_2/n^2+\dots$.
        The first coefficients $\eta_1,\Dl_1,\ld{1}$ for the special case of
        dimension $D=2\mu=3$ have been calculated in
        \ci{Gat}. The coefficient $\eta_1$ for arbitrary dimension
        $D=2\mu$ have been calculated in \ci{Muta},
        $\Delta_1$ in \ci{Zinn} and $\lambda_1$, $\eta_2$
        in \ci{Gr}. In \ci{TMF} coefficients
        $\Dl_2,\ld{2}$ and $\eta_3$ have been calculated.

        The expression for $\eta_3$ contains function $I(\mu)$ which
        cannot be expressed explicitly through the gamma-function $\Gm(z)$
        and its derivatives.  The function $I(\mu)$ is expressed through
        the self-energy massless diagram showed in Fig.1 on which a line
        with index $a$ means a massless propagator $|x-y|^{-2a}$. According
        to its dimension the diagram is a simple line with index
        $\mu-1+\Dl$ multiplied by the coefficient $\Pi(\mu,\Dl)$ which is
        the ``value'' of the diagram.  The quantity $I(\mu)$ is defined
        from $\Pi(\mu,\Dl)$ by the relation
        $I(\mu)=d\ln\Pi(\mu,\Dl)/d\Dl|_{\Dl=0}$ (for $\Dl=0$ the diagram
        is calculated exactly:  $\Pi(\mu,0)=3\pi^{2\mu}(\psi^\prime(\mu-1)
        -\psi^\prime(1)) \Gm(\mu-1) \Gm(2-\mu)/\Gm(2\mu-2)$).  In \ci{Vas}
        the value of $I(\mu)$ for the dimension $D=2\mu=3$ and the first
        terms of \eps\ expansions around $D=4$ are given:  $I(3/2)=
        3\psi^{\prime\prime} (1/2)/2 \pi^2+2\ln 2$, $I(2-\veps)=O(\veps)$
        and in \ci{Vas,Wegner} the \veps\ expansion around $D=2$ was
        obtained
\be
        {\displaystyle
        I(1+\veps)=-\frac{2}{3\veps}+\frac{2\zeta(3)
        \veps^2}{3}- \zeta(4)\veps^3+\frac{13\zeta(5)\veps^4}{3}+\dots}
\lab{11d}
\ee
        An alternative way of derivation of \veps\ expansion for $I(\mu)$
        is presented in the Appendix.

	Using $1/n$ expansion of critical exponents
	and some additional information about the explicit $n$
        dependence of RG functions one can derive high loop
	approximation for RG functions.  This additional information
        usually includes the knowledge of exact expressions of the
        RG functions for some special values of $n$.

\section{Five loop approximation for the Gross Neveu and the $\sg$
	models.}

        As have been noted in \ci{Ros} the case $n=1$ ($N=1/2$) corresponds
        to Majorana spinor and, hence, there are no four-fermion
        interaction term in \re{1}, \re{2}, \re{10}, \re{11} and we have
        for model \re{2}
\be
        Z_0=0, Z_1=Z_2=1
\lab{18}
\ee
        and $Z_3$ is undefined as coefficient of zero.

        Note that though the interaction $(\psb\psi)^2$ does not
        exists at $n=1$ but $\sg\psb\psi$ does and for \re{11} we have
\be
        \bar{Z}_1=1\ ,\ \ \bar{Z}^{-1}_2=\bar{Z}^{-1}_3=\bar{Z}^{-1}_4=
        1+\frac{\ld{}^2}{4\pi\veps}\ ,\ \ \bar{Z}_0=-\frac{\ld{}^2}{32\pi
        \veps}\bar{Z}_2\ ,\ \
        \bar{Z}_6=\frac{\ld{}}{4\pi\veps}\bar{Z}_2
\lab{14}
\ee
        but again we can say anything about the value of $\bar{Z}_5$
        at $n=1$.

        From above considerations we conclude that for the model \re{2}
        $\gm_\psi(g)|_{n=1}=\gm_m(g)|_{n=1}=0$ and, hence, these RG
        functions should contain the factor $(n-1)$ at each order of
        perturbation theory. But we cannot say the same about the $\bt$
        function.

        The case $n=2$ corresponds to $N=1,D=2$. As have been noted
        very early the $U(1)$ Thirring and $N=1$ Gross Neveu model
	are equivalent in the two dimensions. The Thirring model in
	the two dimensions is known to be exactly soluble and it is
        easy to derive the RG functions ($\veps=0,N=1$):
\be
        \beta(u)=0\ ,\ \ \gamma_\psi(u)=\frac{u^2}{1+2u}\ ,\ \
        u\equiv\frac{g}{4\pi}\ .
\lab{15}
\ee
        But from \ci{Gr} it is known the following three loop
        approximation for RG functions in the model \re{2}:
\be
        \begin{array}{l}
        {\gps}=u^2(n-1)-u^3(n-1)(n-2)+{\rm O}(u^4)\\
        \beta(u)=-2u^2(n-2)+4u^3(n-2)+2u^4(n-2)(n-7)+{\rm O}(u^5)
        \end{array}
\lab{16d}
\ee
        setting $n=2$ in it we obtain
\be
        \begin{array}{l}
        \beta|_{n=2}=0\ ,
        {\gps}|_{n=2}=u^2+{\rm O}(u^4)\ .
        \end{array}
\lab{17d}
\ee
        From the above results we can say that in the framework of
        $2+\veps$ expansion the Thirring and the Gross Neveu models are
        not completely equivalent. Hence, there are no reason to believe
        that higher loop contributions of $\gamma_\psi$ should contain
        the factor $n-2$.

        Using now this information and results of \ci{TMF} it is possible
        to reconstruct RG functions of the Gross Neveu model in five loop
        approximation (up to some unknown coefficients) by expanding the
        critical exponents in both \veps\ and \nn\ and comparing then
        the corresponding coefficients.  The result is the following:
\begin{eqnarray}
        \gamma_\psi(u)&=&u^2(n-1)-u^3(n-1)(n-2)+
        u^4(n-1)[n^2-7n+7]+\nonumber\\
        &&+u^5\frac{(n-1)}{3}[3n^3(-2\zeta(3)-1)+2n^2(-3\zeta(3)+17)
        \nonumber\\
        &&\qquad\qquad+12n(-11\zeta(3)-10)+\eta_{50}]
\end{eqnarray}

\begin{eqnarray}
        \beta(u)&=&2u\veps -2u^2(n-2)+4u^3(n-2)
        +2u^4(n-2)(n-7)+\nonumber\\
        &&+\frac{4}{3}u^5(n-2)[-n^2-n(66\zeta(3)+19)+\beta_{40}]+
        \nonumber\\
        &&+\frac{1}{3}u^6(n-2)[3n^3(-2\zeta(3)+1)
        +n^2(261\zeta(4)-168\zeta(3)-83)+\nonumber\\
        &&\qquad\qquad+n\beta_{51}+\beta_{50}]
\end{eqnarray}

\begin{eqnarray}
        \gamma_m(u)&=&2u(n-1)-2u^2(n-1)-2u^3(n-2)(2n-3)+\nonumber\\
        &&+2u^4\frac{(n-1)}{3}[n^2(-6\zeta(3)+5)+n(90\zeta(3)+23)+m_{40}]
        +\nonumber\\
        &&+u^5\frac{(n-1)}{3}[3n^3(-6\zeta(3)-5)
        +n^2(-81\zeta(4)+264\zeta(3)+169)+\nonumber\\
        &&\qquad\qquad+nm_{51}+m_{50}]
\end{eqnarray}
        Where
\begin{equation}
        \gamma_m=-\gamma[\bar\psi\psi]=2\eps-\gamma_\sigma
\end{equation}
        is the anomalous dimension of the mass and $\eta_{50},
        \beta_{40},\beta_{51},\beta_{50},m_{40},m_{51},m_{50}$ are
        unknown coefficients. In \ci{Gr} the expression $(n-2)(n-5)=
        n^2-7n+10$ for the four loop contribution of $\gamma_\psi$ have
        been received instead $n^2-7n+7$. This result have been derived
        in \ci{Gr} using expressions for $\eta_1,\eta_2$ and due to
        the supposition that multiplier $(n-2)$ should be extracted. We
        checked our result (7 instead 10) by direct calculation of 30
        relevant four loop diagrams in the MS scheme.  So the supposition
        of extracting of multiplier $n-2$ to all orders in $\gamma_\psi$ RG
        function is wrong.

        Using a similar method and results of \ci{SEP} it is possible to
        derive the following expressions for RG functions up to five
        loops in  the non-linear \sg\ model:
\begin{eqnarray}
        \gamma_\varphi&=&\frac{u}{2}-u^2\frac{(n-2)}{2}
        +u^3\frac{(n-2)(2n-5)}{8}+\nonumber\\
        &&+u^4\frac{(n-2)}{24}[-3n^2+2n+14+3(3-n)(n+5)\zeta(3)]+\nonumber\\
        &&+u^5\frac{(n-2)}{192}[6n^3(3\zeta(4)+4\zeta(3)+2)
        -n^2(54\zeta(3)+5)\nonumber\\
        &&-n(315\zeta(4)+1764\zeta(3)+307)+a]
\end{eqnarray}
\begin{eqnarray}
        \beta&=&2\veps u-u^2(n-2)-u^3(n-2)
        -u^4\frac{(n-2)(n+2)}{4}-\nonumber\\
        &&-u^5\frac{(n-2)}{12}[-n^2+22n-34+18(n-3)\zeta(3)]+\nonumber\\
        &&+u^6\frac{(n-2)}{96}[3n^3(2\zeta(3)-1)
        +n^2(81\zeta(4)+144\zeta(3)-11)+bn+c]
\end{eqnarray}
        where $a,b,c$ are unknown coefficients.


\newpage
\renewcommand{\theequation}{A.\arabic{equation}}

\setcounter{equation}{0}
\section*{Appendix}
        In this appendix we give the main steps of derivation of recurrent
        relation for function $I(\mu)$. This method allow to obtain the
        expansion \re{11d} without the procedure of summing infinite series
        of the Gegenbauer polinomials \ci{Wegner}. The definition of
        $I(\mu)$ is
\be
        I(\mu)\equiv\left.\frac{{\rm d}}{{\rm d}\Dl}\ln
        \Pi(\mu,\Dl) \right|_{\Dl=0}\ ,
\lab{A.1}
\ee
        where (see Fig.1)
\be
        \int\!{\rm d}^Dx_3{\rm d}^Dx_4\ x_{13}^{-2}
        x_{14}^{-2} x_{34}^{-2(\mu-1+\Dl)} x_{23}^{-2(\mu-1)}
        x_{24}^{-2(\mu-1)}=\Pi(\mu,\Dl)\cdot x_{12}^{-2(\mu-1+\Dl)}\ .
\lab{A.2}
\ee
        Using transformation
        ``$\leftarrow$'' from \ci{SEP} the function $\Pi(\mu,\Dl)$ can be
        transformed to
\be
        \Pi(\mu,\Dl)=c\cdot F_\mu(2\mu-3+\Dl)\ ,
\lab{A.3}
\ee
        where
\be
        \int\!{\rm d}^Dx_3{\rm d}^Dx_4\ x_{13}^{-2} x_{14}^{-2}
        x_{34}^{-2\al} x_{23}^{-2} x_{24}^{-2}=
        F_\mu(\al)\cdot x_{12}^{-2(\al+4-2\mu)}
\lab{A.4}
\ee
        and
\be
        c=\frac{\Gamma(\mu-1+\Dl)\Gamma(\mu-1-\Dl)}{
        \Gamma(\mu-1)\Gamma(\mu-1)\Gamma(1+\Dl)\Gamma(1-\Dl)}=1+
        {\rm O}(\Dl^2)\ .
\lab{A.5}
\ee
        From \re{A.5} it is obvious that the coefficient $c$ is not
        contributed to $I(\mu)$ and we have
\be
        I(\mu)\equiv\left.\frac{{\rm d}}{{\rm d}\Dl}\ln F_\mu(2\mu-3+\Dl)
        \right|_{\Dl=0}\ .
\lab{A.6}
\ee
        Using the Gram determinant method \ci{Pis} one can obtain the
        following relation with shifted dimension of the space
\be
        -F_\mu(\al)-F_\mu(\al+1)+ F_\mu^{(1)}(\al)=
        \frac{2(2\mu-1)(\mu-1)}{\pi^2}F_{\mu+1}(\al+2)\ ,
\lab{A.7}
\ee
        where $F_\mu^{(1)}(\al)$ is a sum of integrals which can be
        integrated using the chain rule. This sum is
\ba
        F_\mu^{(1)}(\al)&=&-4\pi^D[\al^2+\al(-3\mu+5)+\mu^2-5\mu+5]
        \times\nonumber\\
        &&\times\frac{\Gamma(\mu-1)\Gamma(\mu-1)\Gamma(\mu-2-\al)
        \Gamma(\al+3-2\mu)}{\Gamma(\al+2)\Gamma(3\mu-3-\al)}\ .
\lab{A.8}
\ea
        On other hand, combining two recurrent relations \ci{SEP} for the
        master diagram one can derive the following relation for
        $F_\mu(\al)$:
\be
        F_\mu(\al+1)=-\frac{(2\mu-3-\al)}{(\mu-2-\al)}F_\mu(\al)+
        F_\mu^{(2)}(\al)\ ,
\lab{A.9}
\ee
        where $F_\mu^{(2)}(\al)$ is again a sum of easily integrated
        diagrams:
\be
        F_\mu^{(2)}(\al)=-\frac{2\pi^D(3\mu-5-2\al)}{(2\mu-3-\al)
        (\al+2-\mu)^2}
        \frac{\Gamma(\mu-1)\Gamma(\mu-1)\Gamma(\mu-1-\al)
        \Gamma(\al+4-2\mu)}{\Gamma(\al+1)\Gamma(3\mu-4-\al)}\ .
\lab{A.10}
\ee
        From \re{A.7} and \re{A.9} we get
\be
        F_\mu(\al)=\frac{2(2\mu-1)(\mu-\al-2)}{\pi^2}F_{\mu+1}(\al+2)
        +R_\mu(\al)\ ,
\lab{A.11}
\ee
\ba
        R_\mu(\al)&=&\frac{(\mu-2-\al)}{(\mu-1)}[F_\mu^{(2)}(\al)-
        F_\mu^{(1)}(\al)]\nonumber\\
        &=&-2\pi^D(\al+\mu)
        \frac{\Gamma(\mu-1)\Gamma(\mu-1)\Gamma(\mu-2-\al)
        \Gamma(\al+4-2\mu)}{\Gamma(\al+2)\Gamma(3\mu-3-\al)}\ .
\lab{A.12}
\ea
        Substituting now $\al=2\mu-3+\Dl$ and accounting that at $\Dl=0$
        \re{A.2} can be integrated exactly (see, for example, \ci{SEP}):
\be
        \Pi(\mu,0)=F_\mu(2\mu-3)=3\pi^D\frac{\Gamma(\mu-1)\Gamma(2-\mu)}{
        \Gamma(2\mu-2)}(\psi^\prime(\mu-1)-\psi^\prime(1))\ ,
\lab{A.13}
\ee
        we derive the following relation for $I(\mu)$
\be
        I(\mu)=\frac{1}{(\mu-1)^2 A}\left[B+\frac{1}{6(\mu-1)}\right]+
        \left[1+\frac{1}{(\mu-1)^2
        A}\right]\left[I(\mu+1)+\frac{1}{\mu-1}\right]\ ,
\lab{A.14}
\ee
        where
\be
        A=\psi^\prime(1)-\psi^\prime(\mu-1)\ ,\ \
        B=\psi(2\mu-2)+\psi(2-\mu)-\psi(\mu-1)-\psi(1)\ .
\lab{A.15}
\ee
        Due to the property $I(2+\veps)={\rm O}(\veps)$ \ci{SEP,TMF} one can
        obtain the expansion of $I(1+\veps)$ ($D=2+2\veps$) in $\veps$ up to
        order $\veps^3$:
\be
        I(1+\veps)=-\frac{2}{3\veps}+\frac{2\zeta(3)}{3}\veps^2-\zeta(4)
        \veps^3 + {\rm O}(\veps^4)\ .
\lab{A.16}
\ee



\newpage


\unitlength=2.00mm
\special{em:linewidth 0.6pt}
\linethickness{1.0pt}
\begin{center}
\begin{picture}(70.00,20.00)(0,0)
        \put(1.00,10.00){\circle*{1.00}}
        \put(21.00,10.00){\circle*{1.00}}
        \emline{1.00}{10.00}{1}{11.00}{17.00}{2}
        \emline{1.00}{10.00}{3}{11.00}{3.00}{4}
        \emline{11.00}{17.00}{5}{21.00}{10.00}{6}
        \emline{11.00}{3.00}{7}{21.00}{10.00}{8}
        \emline{11.00}{3.00}{9}{11.00}{17.00}{10}
        \put(6.00,15.50){\makebox(0,0)[cc]{${}_1$}}
        \put(6.00,4.50){\makebox(0,0)[cc]{${}_1$}}
        \put(17.00,4.00){\makebox(0,0)[cc]{${}_{\mu-1}$}}
        \put(17.00,16.00){\makebox(0,0)[cc]{${}_{\mu-1}$}}
        \put(15.00,10.00){\makebox(0,0)[cc]{${}_{\mu-1+\Dl}$}}
        \put(24.00,10.00){\makebox(0,0)[cc]{$\equiv$}}
        \put(30.00,10.00){\makebox(0,0)[cc]{$\Pi(\mu,\Dl)$}}
        \put(34.50,10.00){\makebox(0,0)[cc]{$\cdot$}}
        \put(36.00,10.00){\circle*{1.00}}
        \put(46.00,10.00){\circle*{1.00}}
        \emline{36.00}{10.00}{11}{46.00}{10.00}{12}
        \put(41.00,8.00){\makebox(0,0)[cc]{${}_{\mu-1+\Dl}$}}
        \put(48.00,9.50){\makebox(0,0)[cb]{,}}
        \put(62.00,10.00){\makebox(0,0)[cc]{$I(\mu)\equiv\left.
        {\displaystyle {d\over d\Dl} } \ln \Pi(\mu,\Dl) \right|_{\Dl=0}$}}
        \put(37.00,-1.00){\makebox(0,0)[cc]{\large Fig.1}}
\end{picture}
\end{center}


\newpage

\begin{thebibliography}{99}

\bibitem{Gat} G.Gat, A.Kovner, B.Rosenstein, Phys.Lett.,
		\underline{240B}, 158, 1990.

\bibitem{Muta} S. Hikami, T. Muta, Prog. Theor. Phys. {\bf 57} (1977),
         785.

\bibitem{Zinn} J.Zinn-Justin, Nucl.Phys., \underline{B367}, 105,
	1991.

\bibitem{Gr} J.A.Gracey, Int.J.Mod.Phys., \underline{A6}, 395, 1991, \\
             J.A.Gracey, Int.J.Mod.Phys., \underline{A6}, 2755, 1991.

\bibitem{TMF} A.N.Vasil'ev, S.E.Derkachov, N.A.Kivel, A.S.Stepanenko,
               Teor. Mat. Fiz., v.92, N3, 486--498 (1992); v.94, N2,
               179--193 (1993); preprint T93/016, Saclay.\\
	      A.N.Vasil'ev, A.S.Stepanenko,
              ``$1/n$--expansion in the Gross--Neveu Model: Computation
               of $1/\nu$ Exponent at Order $1/n^2$ with Conformal Bootstrap
              Method'', submitted to publication to Teor. Mat. Fiz.

\bibitem{Vas} A.N.Vasil'ev, Yu.M.Pis'mak,Yu.R.Khonkonen, Teor. Mat.
		Fiz., {\bf 50}, 195 (1982).

\bibitem{Wegner} W. Bernreuther, F. Wegner, Phys. Rev. Lett. {\bf 57}
         (1986), 1383.

\bibitem{Ros} C.Luperini, P.Rossi, Ann.Phys., 212, 371--401, 1991.

\bibitem{SEP} A.N.Vasil'ev, Yu.M.Pis'mak,Yu.R.Khonkonen, Teor. Mat. Fiz.,
                {\bf 46}, 157 (1980). \\
              A.N.Vasil'ev, Yu.M.Pis'mak,Yu.R.Khonkonen, Teor. Mat. Fiz.,
                {\bf 47}, 291 (1981).

\bibitem{Pis} S.E.Derkachov, J.Honkonen, Yu.M.Pis'mak. J. Phys. A: Math.
        Gen. {\bf 23} (1990) 5563--5576.

\end{thebibliography}
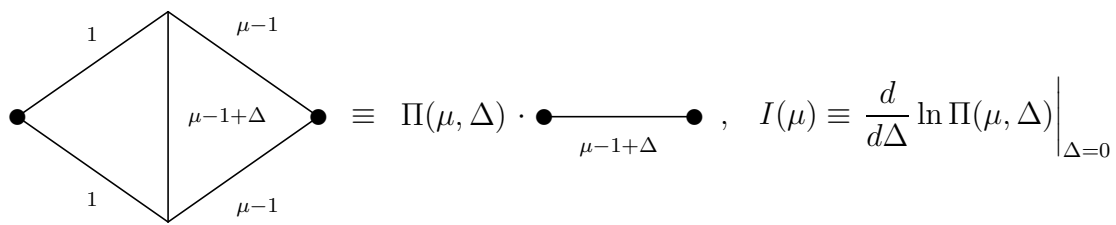
\end{document}